\documentclass[aps,prb,reprint,showpacs,superscriptaddress]{revtex4-1}
\usepackage{graphicx} 
\usepackage{bm}

\begin{document}

\title{Effect of the change in the interface structure of Pd(100)/SrTiO$_3$ for quantum-well induced ferromagnetism}

\author{Shunsuke Sakuragi}
\email[Electronic address: ]{sakuragi@az.appi.keio.ac.jp}
\affiliation{Department of Applied Physics and Physico-Informatics, Faculty of Science and Technology, Keio University, Hiyoshi, Yokohama 223-0061, Japan}

\author{Tomoyuki Ogawa}
\affiliation{Department of Electronic Engineering, Graduate School of Engineering, Tohoku University, Aza-Aoba, Aramaki, Aoba-ku, Sendai 980-8579, Japan}

\author{Tetsuya Sato}
\affiliation{Department of Applied Physics and Physico-Informatics, Faculty of Science and Technology, Keio University, Hiyoshi, Yokohama 223-0061, Japan}

\date{\today}

\begin{abstract}
Pd(100) ultrathin films show ferromagnetism induced by the confinement of electrons in the film, i.e., the quantum-well mechanism. In this study, we investigate the effect of the change in the interface structure between a Pd film and SrTiO$_3$ substrate on quantum-well induced ferromagnetism using the structural phase transition of SrTiO$_3$. During repeated measurement of temperature-dependent magnetization of the Pd/SrTiO$_3$ system, cracks were induced in the Pd overlayer near the interface region by the structural phase transition of SrTiO$_3$, thereby changing the film-thickness dependence of the magnetic moment. This is explained by the concept that as the magnetic moment in Pd(100) changed, so too did the thickness of the quantum-well. In addition, we observed that the ferromagnetism in the Pd(100) disappeared with the accumulation of cracks due to the repetition of the temperature cycle through the phase-transition temperature. This suggests that lowering the crystallinity of the interface structure by producing a large number of cracks has a negative effect on quantum-well induced ferromagnetism. 
\end{abstract}


\maketitle

\section{INTRODUCTION}
Nanosized metal has different electric states due to the restriction of the movement of electrons near the Fermi energy $\epsilon _F$. 
In film form, these are called the quantum-well states (QWs), and the movement of electrons is confined between the surface and the interface of the film \cite{chiang}. 
Elementary quantum mechanics describes how the electric states of QWs are modulated periodically, depending on the film thickness. As the physical properties of metal are determined by the electrons near the $\epsilon _F$, QWs modulate the physical properties of metal materials (e.g., electric conduction\cite{DendouPRL}, superconducting transition temperature\cite{GuoScience}, and magnetic properties\cite{WeberPRL, ortegaPRB, BauerPRB, DabrowskiPRL}). 
In Pd, which is a nonmagnetic 4$d$ transition metal, QWs induce ferromagnetism in an oscillatory manner, depending on film thickness\cite{SakuragiPRB}. 
This is explained in terms of the increase of $D(\epsilon _F)$, which originates from QWs in Stoner's criterion $ID(\epsilon _F)>1$ for ferromagnetism in metal, where $I$ is the exchange integral, and $D$ is the density of states\cite{stoner}. 
Based on this, nonmagnetic-ferromagnetic switching\cite{sunPRB, aiharaJAP} is expected through the modulation of QWs using an electric field.

In QWs, electrons near the $\epsilon _F$ are reflected at the surface and the interface of the film. 
Thus, the scattering of electrons at the interface is significant for the formation of QWs. 
In Pd(100) ultrathin films obtained by epitaxial growth on SrTiO$_3$ (STO) substrates, a steep Pd/STO interface structure suitable for the QW-induced ferromagnetism was formed\cite{PhysProc}. 
A change in the interface structure of Pd/STO is expected as the temperature cycles between room- and low-temperature, because the cubic to tetragonal structural phase transition occurs in STO at the critical temperature $T_a \sim 105$ K\cite{wang}. 
This can change the properties of QWs. 
If such a modulating mechanism exists, it should be possible to observe the magnetic phase transition and/or change in the spontaneous magnetization of Pd/STO by means of temperature-dependent magnetic measurements.

In this paper, we studied the temperature-dependent magnetic measurement of Pd(100) ultrathin films on STO substrates to investigate the relationship between changes in the interface structure due to the phase transition of STO and the change in QW-induced ferromagnetism. 
We observed the jump in magnetization due to the phase transition of STO, and the disappearance of ferromagnetism after temperature-cycle repetition. 
This was thought to be because the density of a few layers in the Pd film decreased near the Pd/STO interface to lower the stability of the QW-induced ferromagnetism. 
The formation of a low-density layer lead to a modulation in the effective thickness of QWs, and thus the change in the magnetic moment depending on the film thickness. 
As the low-density layer increased, the crystallinity of Pd layers near the Pd/STO interface was significantly lowered, and finally, the ferromagnetism of Pd disappeared.

\section{EXPERIMENT}
We grew atomically flat Pd(100) ultrathin films on buffered HF-treated STO(100) substrates (SHINKOSHA Co., Ltd.\cite{KawasakiScience}) using the three-step growth method\cite{wagnerJAP} in an ultrahigh vacuum chamber with base pressure below $1\times 10 ^{-9}$ Torr. 
After deposition, we encapsulated the sample in a quartz tube without exposing it to air to prevent gas adsorption. 
We could thus measure the magnetization of a sample with a clean surface. 
The X-ray crystal truncation rod scattering of Pd/STO that indicates epitaxial growth was observed in our samples\cite{PhysProc}. 
The details of sample preparation have been described elsewhere\cite{SakuragiPRB}. 
The magnetization measurements were performed using a superconducting quantum interference device magnetometer (Quantum Design MPMS-XL and MPMS-3). 
The temperature dependence of the magnetic moment was measured in a temperature range between 10 and 300 K, and in a magnetic field of 1.5 T.

\section{RESULTS}
\subsection{Characterization of the magnetic properties}
We prepared two samples and measured the magnetization at 300 K. 
We then took the Pd/STO sample out of the quartz tube and measured the thickness and surface roughness of the film using X-ray reflectivity (XRR) after the all magnetization measurement. 
Table 1 shows the saturation magnetic moment per atom ($m_s$) of samples A and B, estimated from the measurements of magnetization and XRR. 
The values of $m_s$ were consistent with our previously reported data of the thickness dependence of $m_s$\cite{SakuragiPRB}. 
In addition, Table 1 shows a surface roughness of 2 and 1 monolayers (MLs) of Pd in samples A and B, respectively. 
This suggested that sample A had a wider distribution of thickness in the film compared to sample B. 

After cooling the sample, we measured the temperature dependence of the magnetic moment in the heating process [Figs. 1(a) and (b)], where the total magnetic moment of the Pd(100) film, STO substrate, and quartz sample tube were measured. 
The sharp change in the magnetic moment was observed at a temperature between 120 K and 150 K, which was higher than the phase transition temperature $T_a$ of STO in both samples. 
Such a change was not observed in the blank sample, which contained only STO and the quartz tube, and thus this behavior was caused by the change in the magnetic moment of Pd(100). 
The change in magnetic moment was observed as a sudden decrease in sample A, and a weak increase in sample B. 
This suggests that the singular behavior of the magnetic moment in Pd/STO was dependent on the surface roughness.

\begin{table}
\caption{Properties of the Pd(100) films on the STO substrate. The $m_s$ was measured at 300 K and the thickness and surface roughness were measured using XRR at room temperature.}
\begin{ruledtabular}
\begin{tabular}{cccc}
 &$m_s$&Thickness&Surface Roughness\\
 &($\mu _B$/atom)&(nm)&(nm)\\
\hline
Sample A&$0.13
$&3.86&0.45\\
Sample B&$0.10
$&3.83&0.26\\
\end{tabular}
\end{ruledtabular}
\end{table}

\begin{figure}
\includegraphics[width=7.5cm]{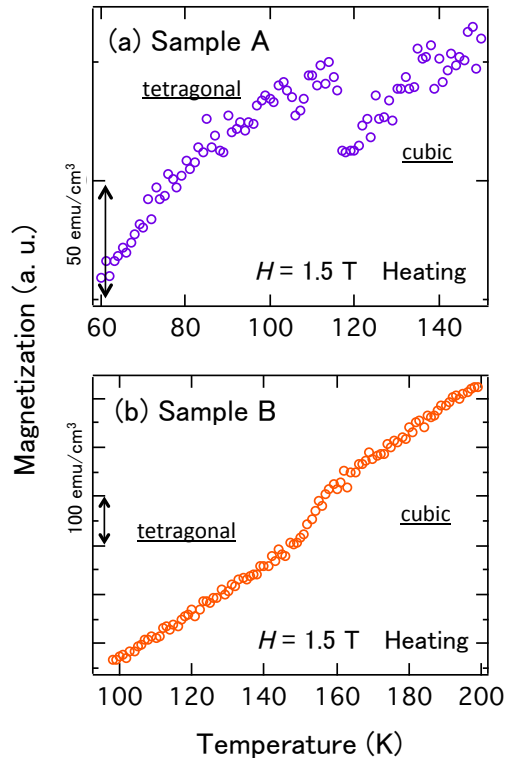}
\caption{Temperature-dependent magnetic moments of samples A (a) and B (b). Jumps of magnetic moment were observed in both samples. The vertical axis shows the magnetic moment, which is divided by the volume of the Pd film. }
\end{figure}

After six repetitions of the temperature cycling between10 K and 300 K in sample B, we measured the temperature-dependent magnetic moments from 300 K to 10 K, and subsequently, from 10 K to 300 K [Fig. 2 (a)]. 
In the cooling process, the sudden change of magnetic moment was observed at $\sim$ 100 K, which was close to the typical temperature of $T_a$ of STO. 
In the heating process, on the other hand, the sudden change was not observed. 
Fig. 2(b) shows the magnetization curves of sample B measured at 300 K before the first and after the seventh temperature cycle. 
Spontaneous magnetization was observed before the temperature cycle, but disappeared after that. 
Thus, the ferromagnetic nature of Pd(100) on STO irreversibly changed to nonmagnetic by the temperature cycling through the phase-transition temperature of STO.

\begin{figure}
\includegraphics[width=7.5cm]{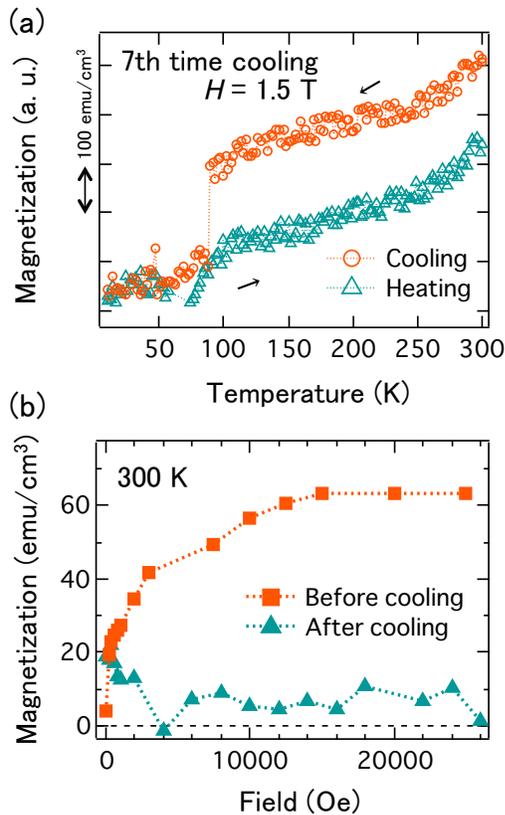}
\caption{Temperature-dependent magnetic moment of sample B measured in the seventh cooling and heating processes (a) and the magnetization curves of sample B measured at 300 K before the first cooling procedure and after the seventh cooling (b). No jump in the temperature-dependent magnetic moment was observed after the seventh cooling (a); (b) shows the disappearance of the spontaneous magnetization of Pd(100) after the seventh cooling. }
\end{figure}

\subsection{Structure analysis before and after cooling}
To investigate the effect of the temperature cycle on the structure of Pd/STO, we obtained the XRR profiles of other Pd/STO samples before and after the cooling procedure under conditions similar to that of sample B [Fig. 3 (a)]. 
After cooling, the increase in the intensity of the peak of fringes and the change in the period of the oscillation were observed in the XRR profile. 
This suggests that the cooling procedure brings about an irreversible change in the structure of the film.

\begin{figure*}
\includegraphics[width=17.8cm]{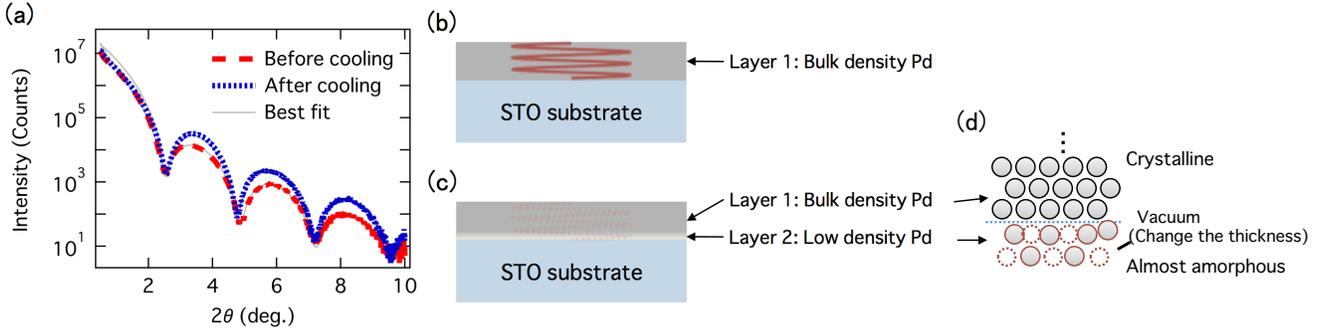}
\caption{XRR profiles of Pd/STO measured before first cooling and after seventh cooling (a). The difference in the profiles indicates the structural modification of Pd by the cooling process. The fitting curves were calculated based on the Pd/STO bilayer before the first cooling (b) and the Pd(bulk density)/Pd(low-density)/STO trilayer model for the sample after the seventh cooling (c). The scheme of the arrangement of atoms near the interface, which was assumed after the seventh cooling, is shown in (d). The cracks of the Pd behave as the vacuum layer, and the accumulation of cracks induced the deterioration of the crystallinity of the Pd layer. }
\end{figure*}

\section{DISCUSSION}
\subsection{Change in the structure by cooling}
We determined structural parameters from the best fit of XRR profiles of Pd/STO obtained before the first and after the seventh cooling to investigate the effect of temperature cycle on film structure. 
In the fitting of the profile obtained before cooling, it was appropriate to assume that the density of the Pd(100) film was uniform [Fig. 3 (b)]; this was not, however, suitable for the profile obtained after cooling. 
Thus, we assumed that the structural modulation of Pd occurred near the Pd/STO interface by the structural phase transition of STO, i.e., the structure of Pd(bulk density)/Pd(low-density)/STO(substrate) [Figs. 3(c) and (d)]. 
Based on the fitting using this assumption, we judged that the density of the 2 MLs of Pd at the interface was reduced by up to $\sim$ 40\% compared to another layer of Pd.

The reduction in the density of Pd layers near the interface can be explained in terms of the generation of cracks in the Pd film near the interface. STO, which exhibits quantum paraelectric behavior, has a cubic structure at room temperature, and changes to a tetragonal structure at the low temperature $T_a$ of $\sim$ 105 K. 
In a single crystal of STO with a tetragonal structure, various domains with $c$-axes are formed in different directions \cite{KaliskyNmat, HonigNmat, MaertenPRL}. 
The domain size is $\sim$ 1 $\mu m$, and the anti-phase domains, which are the boundary of domains with $c$-axes in different directions, induce lattice distortion in the overlayer\cite{EgilmezPRB, LoetzschAPL}. 
Thus, it is expected that the Pd overlayer on the anti-phase domains will be distorted when the cubic-to-tetragonal phase transition occurs at low temperature, and cracks are formed in the Pd layer. 
This causes the irreversible change in the temperature dependence of the magnetic moment. 
Due to the repeated temperature cycles for Pd/STO, the density of the 2 MLs of Pd on the substrate was finally reduced by 40\% due to the accumulation of a large number of cracks. 
Thus, it was theorized that the crystallinity of Pd was intrinsically lowered in the 2 MLs of Pd near the interface by the repetition of temperature cycles [Fig. 3 (d)].

The cubic-to-tetragonal phase transition in STO is caused by the rotation of oxygen octahedra and the expansion of the lattice constant along the $c$-axis direction\cite{wang, ChangPRB}. 
If the rotation of oxygen is clamped by the coupling of the overlayer and STO, the cubic-to-tetragonal phase transition should be affected. 
This phenomenon, which is called the clamping effect, is known as a way to affect the $T_a$ of STO\cite{HeAPL, HePRB, LoetzschAPL}. 
In the present case, the Pd atoms, which were coupled with the oxygen in the Ti-O surface of STO, obstructed the phase transition of STO. 
By this mechanism, the sudden change in magnetic moment of Pd/STO was observed at a temperature between 120 K and 150 K, which was different from the typical temperature of $T_a$ of STO, as shown in Fig. 1. 
After the repetition of temperature cycles, it is considered that the occurrence of cracking near the Pd/STO interface partially removed the coupling of Pd and O. 
This weakened the clamping effect at the Pd/STO interface, and thus the $T_a$ of Pd/STO could be expected to approach the original temperature of $T_a$ of STO\cite{LoetzschAPL}. 
Therefore, the singular behavior in the temperature-dependent magnetic moment of Pd/STO appeared near 100 K after the repetition of temperature cycles, as shown in Fig. 2(a).

\subsection{Effect of the modulation of the interface structure on magnetic property}
Generally, the electronic states in QWs are described by the phase shift quantization rule \cite{chiang}, $2\pi k_z N + (\varphi_s + \varphi_i) = 2\pi j$, where $k_z$ is a discrete wave number in units of $2\pi$/(lattice constant), $N$ is the number of layers, $\varphi_{s, i}$ is a correction term for the scattering phase shift at the surface and interface, and $j$ is an integer quantum number. 
When $k_z$ coincides with the Fermi wave number originating from the $h_5$ band of Pd, the $D(\epsilon _F)$ increases due to the degeneracy of electronic states caused by QWs, and thus ferromagnetism can appear in Pd(100) according to the Stoner criterion\cite{mirbtPRB, niklassonPRB, hongPRB}. 
Therefore, how much $m_s$ is in a Pd(100) ultrathin films is affected by film thickness and the energy potential of the interface, which are the parameters that characterize QWs.

The cracks formed at the Pd/STO interface affected both the thickness of QWs and the electron scattering at the interfaces. 
The existence of $\sim$ 2 MLs cracking layer of Pd, in which the density was reduced by $\sim$ 40\% from the bulk density of Pd by repetition of the temperature cycle, meant that $\sim$ 40\% of the area of the Pd/STO interface became a vacuum region. 
In other words, $\sim$ 60\% of the sample retained the Pd/STO structure, but the remaining part had a Pd/(Vacuum layer)/STO structure. 
Thus, the effective thickness for a QW is 1$\sim$2 MLs smaller than the nominal thickness in part of the sample. 
In addition, the existence of a vacuum layer induced the modulation of the interface potential energy, and thus changed the scattering phase shift of QWs\cite{YoshimatsuPRB}. 
A comparison of the first-principle calculation and the experimental measurement\cite{SakuragiPRB} showed that the thickness necessary to stabilize ferromagnetism in Pd(100) film, which is sandwiched between vacuum layers, is $\sim$ 2MLs thinner than that on the STO substrate. 
Considering this, the effective thickness of QWs also becomes thinner by occurring in the cracks at the Pd/STO interface.

At the initial stage of the temperature cycle, when the accumulation of cracks is low, crystallinity survives in the cracking layer. 
On the other hand, the repetition of temperature cycles decreases crystallinity. 
Fig. 2(b) shows that the spontaneous magnetization disappeared when the crystallinity of the cracking layer of Pd intrinsically deteriorated, as shown in Fig. 3(d), and could not revive in the heating process. 
There are two possible mechanisms to explain this behavior. 
The first is the breakup of QWs by the deterioration of crystallinity near the interface, and the second is the lack of an increase in $D(\epsilon _F)$, which originates from the band broadening of QWs. 
To explain which mechanism is more suitable, it would be necessary to directly observe QW bands using angle-resolved photoemission spectroscopy or detailed theoretical studies. 
Nevertheless, it is obvious that the deterioration of crystallinity has a disadvantageous effect on QW-induced ferromagnetism.

We now consider the difference in the direction of the jump of the magnetic moment due to the structural phase transition of the STO substrate between samples A and B shown in Fig. 1. 
The film thickness of both samples A and B was $\sim$ 3.8 nm, substantially equal. 
Our previous research showed that the spontaneous magnetization of Pd(100)/STO oscillated depending on film thickness, and that one peak in $m_s$ was observed at the thickness of 3.3 nm\cite{SakuragiPRB}. 
Thus, the increase in $m_s$ of samples A and B should occur when the effective thickness of the QWs is reduced by $\sim$ 0.5 nm (2.5 MLs). 
An analysis of Fig. 3(a) shows that the structural phase transition of STO affected the interface Pd layers up to 2 MLs. 
In addition, samples A and B had surface roughnesses (distribution of the thickness) of $\sim$ 2 and $\sim$ 1 MLs (Table 1), respectively. 
Because of the comparatively large distribution of film thickness in sample A, it is suggested that part of sample A had a thickness corresponding to the peak in $m_s$ by the cracking, and that the magnetization was increased at low temperatures, as shown in Fig. 1(a). 
In sample B, which had comparatively small film thickness distribution, the thickness corresponding to the peak in $m_s$ was not achieved anywhere in the sample. 
Thus, the increase in magnetization was not observed at the phase transition to tetragonal structure. 
In addition, the small drop in magnetization of sample B was likely caused by the deterioration of crystallinity in the Pd layers near the interface, although this change was likely hidden by the increase in magnetization in sample A.

In addition, there was the difference in the temperature at which the jumps of magnetic moment occurred between samples A and B shown in Fig. 1. 
This indicated that the strength of the clamping effect depended on the amount of surface roughness. 
The coupling between Pd and STO correlated with the electronic state near $\epsilon _F$ of Pd, which is governed by QWs. 
Thus, the strength of clamping effect should depend on the film thickness through the modulation of QWs, maybe in a manner similar to magnetization. 
This detail will be clarified based on the discussion of the relationship between the changes in structure and electronic state of QWs.

\section{CONCLUSION}
In this study, we systematically degraded the interface structure of Pd/STO using the structural phase transition of the STO substrate. 
When cracks occurred in the Pd overlayer near the substrate, the effective thickness of the QWs was reduced, and the magnetic moment of ferromagnetic Pd(100) induced by QWs changed. 
In addition, the deterioration of crystallinity due to the accumulation of cracks prevented the appearance of QW-induced ferromagnetism. 
This showed that the disturbance of the structure near interface of the overlayer/substrate essentially affected the physical properties induced by QWs. 
These findings increase our understanding of the relationship between the interface structure and the property of QWs. 
As a result, we can discuss the relationship between magnetism and QWs more deeply by changing the interface structure using the structural phase transition of the substrate as a probe.

\begin{acknowledgments}
We thank K. Ueda, N. Awaji, and K. Sakurai on fruitful advice on the analysis of XRR. 
This work was supported by JSPS KAKENHI Grant Numbers \#15J00298 and 15H01998. 
One of the authors (S.S.) also acknowledges a fellowship from the JSPS.
\end{acknowledgments}

\bibliography{ref}
\end{document}